# 3D characterization of the primary Al$_3$Sc phases in an Al-Sc alloy using Synchrotron X-ray tomography and electron microscopy

Yuliang Zhao [a, b*], Weiwen Zhang [b], Billy Koe [c, d], Wenjia Du [c], Mengmeng Wang [e], Weilin Wang [e], Elodie Boller [f], Alexander Rack [f], Zhenzhong Sun [a], Jiawei Mi [c]

[a] Neutron Scattering Technical Engineering Research Centre, School of Mechanical Engineering, Dongguan University of Technology, Dongguan, 523808, P.R. China

[b] National Engineering Research Centre of Near-net-shape Forming for Metallic Materials, South China University of Technology, Guangzhou, 510641, P.R. China

[c] Department of Engineering, University of Hull, East Yorkshire, HU6 7RX, UK

[d] Diamond Light Source, Harwell Science and Innovation Campus, Didcot, Oxfordshire, OX11 0DE, UK

[e] Shanghai Key Laboratory of Advanced High-Temperature Materials and Precision Forming, School of Materials Science and Engineering, Shanghai Jiao Tong University, Shanghai, 200240, P.R. China

[f] ESRF-The European Synchrotron, 71 Avenue des Martyrs, 38000 Grenoble, France

Corresponding author: zhaoyl@dgut.edu.cn (Y. Zhao)

**Abstract**

The three-dimensional structures of the primary Al$_3$Sc particles in an Al-2%Sc master alloy were studied by synchrotron X-ray microtomography, scanning and transmission electron microscopy. The Al$_3$Sc phases were found to be a single cube and a cluster of cubes. The surface area, equivalent diameter of the Al$_3$Sc cubes increased with the increasing of cube volume, but the specific surface area decreases. The primary Al$_3$Sc cubes and Al-matrix have the same crystal orientation, indicating that the Al$_3$Sc phases are the heterogeneous nucleation sites for Al. The experimental results show that α-$Al_2O_3$ are the possible nucleation sites for the $Al_3Sc$ cubes.

Aluminium alloys are widely used in automotive, aerospace and packing industry because of their abundance, high strength-to-weight ratio, good corrosion resistance and excellent recyclability [1]. Grain refinement [2-4] is an effective way to improve both strength and elongation of aluminium alloys. Among all the materials used as Al grain refiners, Sc is one of the most effective elements because it provides the combined effects of (i) grain refinement; (ii) precipitation hardening and (iii) enhanced recrystallization [5-9]. In the Al-Sc master alloys commonly used in industry, $L1_2$ $Al_3Sc$ primary phases formed and act as heterogenous nucleation sites for α-Al because it has a very small mismatch with α-Al [10]. Sc can be also partially replaced by Zr to form $Al_3(ScZr)$ for the purpose of reducing cost and increasing grain refinement efficiency because $Al_3(ScZr)$ has core-shell type nanostructure which have high-strength and excellent thermal-stability [11-12]. Extensive research has been carried out to study and quantify the addition of Sc in different types of Al alloys, and the enhancement of mechanical properties of Al alloys [13-16].

The size, morphology and distribution of $Al_3Sc$ particles have profound effects on the grain refinement efficiency, castability and mechanical properties of the cast ingots [17-18]. Agglomeration of primary particles in the master alloys leads to the reduction of grain refinement efficiency and has the detrimental effects on the final properties [19]. A full understanding of the particle size, 3D morphology and distribution is essential for developing new strategies to improve the grain refinement and properties of Al alloys. Current studies of the morphology of the $Al_3Sc$ phases are performed mainly using optical microscope (OM) or scanning electron microscope (SEM), which cannot obtain the true 3D information. For example, Li *et al.* [15] suggested that 3D morphology of $Al_3(ScZr)$ particles are near-globular structures and located within the a-Al matrix or grain boundary. Kastner *et al.* [20] have observed that primary $Al_3(ScZr)$ particles exhibits equi-axes shapes.

Using high-brilliance synchrotron X-ray, it is possible to obtain 3D information for the phases (particles) with the size above ~ 1 micrometre [21]. However, so far, very limited literatures have reported the use of synchrotron radiation X-ray tomography to characterise $Al_3Sc$ phases in Al alloys.

Thus, in this work, we use synchrotron X-ray tomography to study the size, morphology and distribution of the primary $Al_3Sc$ phases in an Al-2%Sc alloy. In addition, scanning and transmission electron microscopy plus electron backscatter diffraction (EBSD) were used complementarily to further quantify the possible nucleation site for the primary $Al_3Sc$ phases and the orientation relationship between primary $Al_3Sc$ phases and Al matrix.

Al-2.0 wt.% Sc master alloy was used in this study. For conventional 2D microstructure characterisation, a Nova SEM 430 Field Emission Gun scanning electron microscope (SEM) equipped with an Oxford energy-dispersive X-ray spectrometer were used. The samples were polished by Ion Beam Milling System Leica EM TIC 3X machine at 5 kV for 2 h and 3 kV for 0.5 h. Some of the sample was deeply etched by the mixture of high-purity iodine and methanol solution (10 g iodine per 100 mL methanol) at room temperature for approximately 4–5 h to dissolve the Al matrix [22]. The sample was cut in a GAIA3 GMU Model 2016 FIB-SEM microscope operated at 5 kV. Crystal orientation and grain size map were obtained using the HKL

Channel 5 EBSD acquisition system. Owing to the similar crystal structure of primary $Al_3Sc$ particle and α-Al matrix, it is impossible to distinguish the two phases using conventional methods. We used the True Phase model embedded in the HKL Channel 5 to distinguish the two phases. For the TEM thin foils, the possible nucleate particles together with the primary $Al_3Sc$ particles were cut and lifted by a tungsten manipulator and transferred to an Omni lift-out cooper grid in a FEI Quanta 3D FEG dual beam microscope. After the transfer, the relatively thick foils on the Cu grid were further thinned by the ion beam step by step to around 80 nm. The foils were then examined in a JEOL 2100F TEM operated at 200 kV.

Synchrotron X-ray tomography experiments were performed at the ID 19 beamline, at the European Synchrotron Radiation Facility (ESRF), in France. The detailed sample parameters used are like the Refs. [23-25]. The imaging system consists of a 25 μm LuAG: Ce scintillator coupled to a white-beam compatible microscope with a 10 × magnification and a high-speed CMOS camera (PCO. DIMAX). The effective pixel size is 1.2 μm and the tomogram scan time is 1 s. For each scan, 1000 projections were acquired over 180° of sample rotation. Tomographic reconstructions were performed on the ESRF cluster using the inhouse developed software-package PyHST_2. Fig. s1a shows the typical image processing procedures and methods employed for 3D microstructure determination. As showed in Fig. s1b, the grey block phases were identified as $Al_3Sc$. Both X-ray and SEM images show that the primary $Al_3Sc$ particles exhibit square and triangular shapes in 2D. Open source image processing software, Image J [26] was used to adjust the contrast between the different phases. Then, the 3D bilateral filter was applied to the tomography datasets to increase the contrast and reduce noise. Finally, the primary $Al_3Sc$ particles and Al dendrites, were manually trimmed for the different material phases. 3D segmentation and feature rendering were performed using Avizo Lite v9.0.1 (VSG, France). Typically, a sub-volume of $200^3$ voxels with a voxel size of $(1.2\ \mu m)^3$ was chosen for further analyses.

The two parameters were used to characterize the surface curvature, the mean curvature $H$, Gaussian curvature $K$ [27] is defined as:

$$H = 0.5 * \left(\frac{1}{R_1} + \frac{1}{R_2}\right) \qquad (1)$$

$$K = \left(\frac{1}{R_1 * R_2}\right) \qquad (2)$$

where $R_1$ and $R_2$ are the two principal radii of curves respectively. Local curvature is an important geometrical parameter for the interface between two phases (dendrites or intermetallic) formed during the solidification processes, influencing the diffusion of solutes and therefore the final morphology of the phases. We used the skeletonization function available in Avizo® [28] to peel off the 3D structures of the primary $Al_3Sc$ particles down to a skeleton (1-voxel thickness) with connecting nodes. The length of the curve between each node, and the number of the connecting nodes can be calculated, and therefore the 3D characteristics of skeleton. The equivalent diameters $D_{eq}$ is defined as average equivalent diameters of the different particles of the considered phase and the volume $V$ represents the whole volume of considered phase [29]

$$D_{eq} = \sqrt[3]{\frac{6V}{\pi}} \qquad (3)$$

The other important parameter specific surface area ($SSA$) is defined as surface area $A$ per unit volume $V$ [30]:

$$SSA = \frac{A}{V} \qquad (4)$$

Fig. 1 shows the 3D rendering of primary cubic $Al_3Sc$ particles in the volume (240 μm × 240 μm × 240 μm). It can be seen that, from Fig. 1a, most of these particles were found to form clusters with several cubic faceted morphologies. Conglomeration of the clusters in the matrix is also observed in the Ref. [10-11, 18]. The primary $Al_3Sc$ particles are firstly formed in the Al-2Sc alloy during solidification, owing to the Sc content is greater than the eutectic composition (0.55% Sc) [8]. Figs. 1b-c show the mean curvature and gaussian curvature of primary $Al_3Sc$ particles, the red colour of cubic edge of particles illustrating the cusped-cubic morphologies of these particles. Fig. 1d shows the distribution of mean curvature and gaussian curvature of primary $Al_3Sc$ particles, indicating both follow gaussian distribution. The distribution peak position (μ) of mean curvature and gaussian curvature of primary $Al_3Sc$ particles are 0.06 and 0.14, respectively, and the standard deviation (σ) of these particles are 00.06 and 0.14, respectively. This indicates that the cubic primary $Al_3Sc$ particles has the cusped edge. Fig. 1e and f show the skeletons of primary $Al_3Sc$ particles and their node length distribution, respectively. The node length of these particles also follows the gaussian distribution and their length are in the range of 5-20 µm. *Much richer and cleaner information, mean and gaussian curvature and skeletons of primary $Al_3Sc$ phases in different view angles are demonstrated in the accompanying video.*

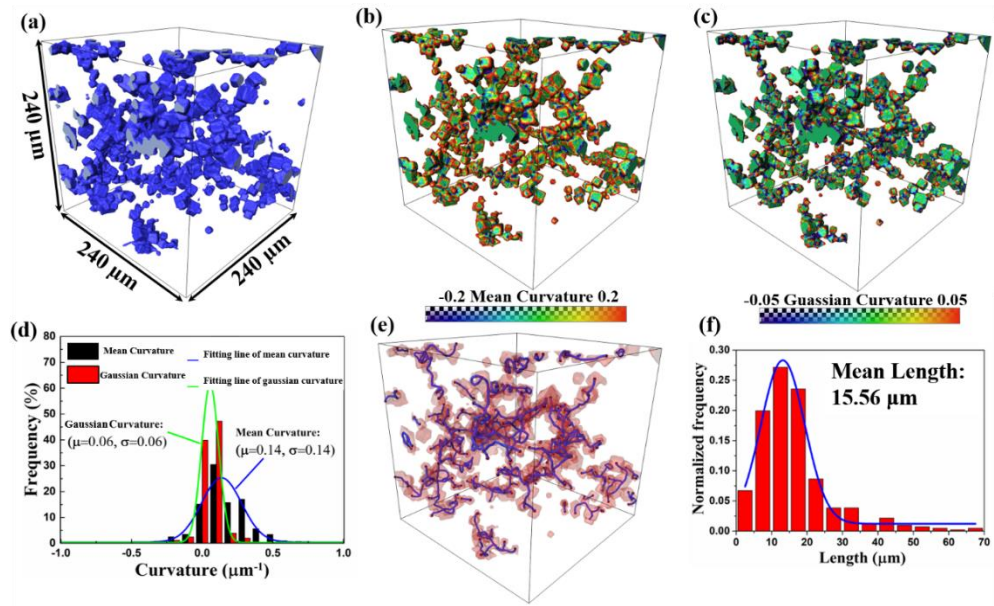

Fig. 1 3D characterization of primary $Al_3Sc$ phases in a volume of 240 μm × 240 μm × 240 μm: (a) 3D reconstructed particles; (b) mean curvature; (c) gaussian curvature; (d) distribution of mean curvature and gaussian curvature; (e) particles using the skeletonization function; (f) statistical analysis of node length.

Typical Al$_3$Sc morphologies found in the Al-2Sc alloy and their statistical analysis are shown in Fig. 2. The cubic primary Al$_3$Sc particles are in contact with each other, which can be clearly seen in both 3D rendered structures and deep-etched SEM images (Fig. 2a). New particles generally nucleate preferentially at the corners or edges of pre-existed particles during the growth of primary Al$_3$Sc particles. This arises because at the particle corners there will be a locally thinner Sc depleted layer ahead of the interface, resulting in higher gradients of constitutional undercooling [31]. Fig. 2b shows the 3D rendered structures and deep-etched SEM images of single Al$_3$Sc particles. From the deep-etched SEM images, the Al$_3$Sc is a typical cubic structure with three perpendicular ledges at the corner. The side length of this single cubic Al$_3$Sc particles is about 25 μm. In some case, there are some holes in the centre-facet of the cubic (Fig. 2c). This similar was also observation in the [17]. This is attributed to the growth of primary Al$_3$Sc particles interface had become locally unstable. Li *et al.* [32] suggested that the central hole of the particle may generate during mechanical grinding and polishing. From the 3D tomography images (Fig. 2c), the central hole is formed during solidification because of the non-destructive nature of synchrotron X-ray tomography. The statistical analysis of Al$_3$Sc cluster and single Al$_3$Sc particle is shown in the Figs. 2d-g. The Fig. 2d shows the volume of Al$_3$Sc cluster is mainly in the range of 5000-80000 μm$^3$ and the insert image shows the volume of single Al$_3$Sc particles is mainly in the range of 1000-4000 μm$^3$. This further confirms that the Al$_3$Sc cluster have a large volume. The next graph (Fig. 2e) shows that volume of primary Al$_3$Sc particles increased with increasing surface area. They follow the linear relationship: Y = -1098 + 2.5*X and single Al$_3$Sc particles are mainly located in the left corner of the figure. This indicating that the Al$_3$Sc with increased surface area is a good grain refiner. In the Fig. 2f, the equivalent diameter of primary Al$_3$Sc particles (including Al$_3$Sc cluster and single Al$_3$Sc particle) is plotted against the volume for the Al-2Sc master alloy. The equivalent diameter of primary Al$_3$Sc particles increased with volume and follow non-linear relationship: Y = 1.24*X$^{3.33}$. Fig. 2g shows the volume primary Al$_3$Sc particles plotted against the specific surface area. Their distribution is almost random. While the larger the volume of primary Al$_3$Sc cluster, the smaller its specific surface area. This means that the increase of volume resulting in the reducing of surface area of Al$_3$Sc cluster and become relatively compacted.

Fig. 3a exhibits the EBSD images of the alloy containing primary Al$_3$Sc particles. Owing to the similar crystal structure of Al and Al$_3$Sc (their mismatch in lattice parameter is 1.34% [8]), it is impossible to distinguish two phases using normal EBSD analysis. The True Phase model embedded in the HKL Channel 5 using to identify the two phases, as shown in Fig. 3b, which is combined with SEM-EDS analysis crystal structure of Al$_3$Sc phase and Al matrix in the alloy. The Kikuchi patterns of Al$_3$Sc and Al phases are shown in the Figs. 3c-d. The calculation of Euler angles ($\varphi_1$= 210.0, $\varphi_2$= 43.2, $\varphi_3$= 82.8) displayed in Fig. 3c for Al$_3$Sc phase, and Euler angles ($\varphi_1$= 80.1, $\varphi_2$= 16.0, $\varphi_3$= 86.0) displayed in Fig. 3d for Al matrix. Mean angular deviation (MAD) is the difference value between the experimental and calculated patterns of the particles calculated by the EBSD software. Usually, the value for MAD is lower than 0.7, which means that the identified particle can be accepted by the calculated patterns of the particle [33]. The Al$_3$Sc phase is formed at 659 °C during a eutectic reaction, has a cubic ordered structure of L1$_2$ type with a = 0.4104 nm. The EBSD results of Al-2Sc alloy is shown in Fig. 3e. This figure shows that the primary Al$_3$Sc particles and Al matrix have the same crystal orientation, which indicating that the Al$_3$Sc particles provide the heterogeneously nucleation site for Al matrix. This result is accordance with previous studies [10, 17-18]. The graph inserted in Fig. 3e displays that the grain size distribution of the alloy is mainly in the range of 20-80 μm and the average grain size is 41.60 μm.

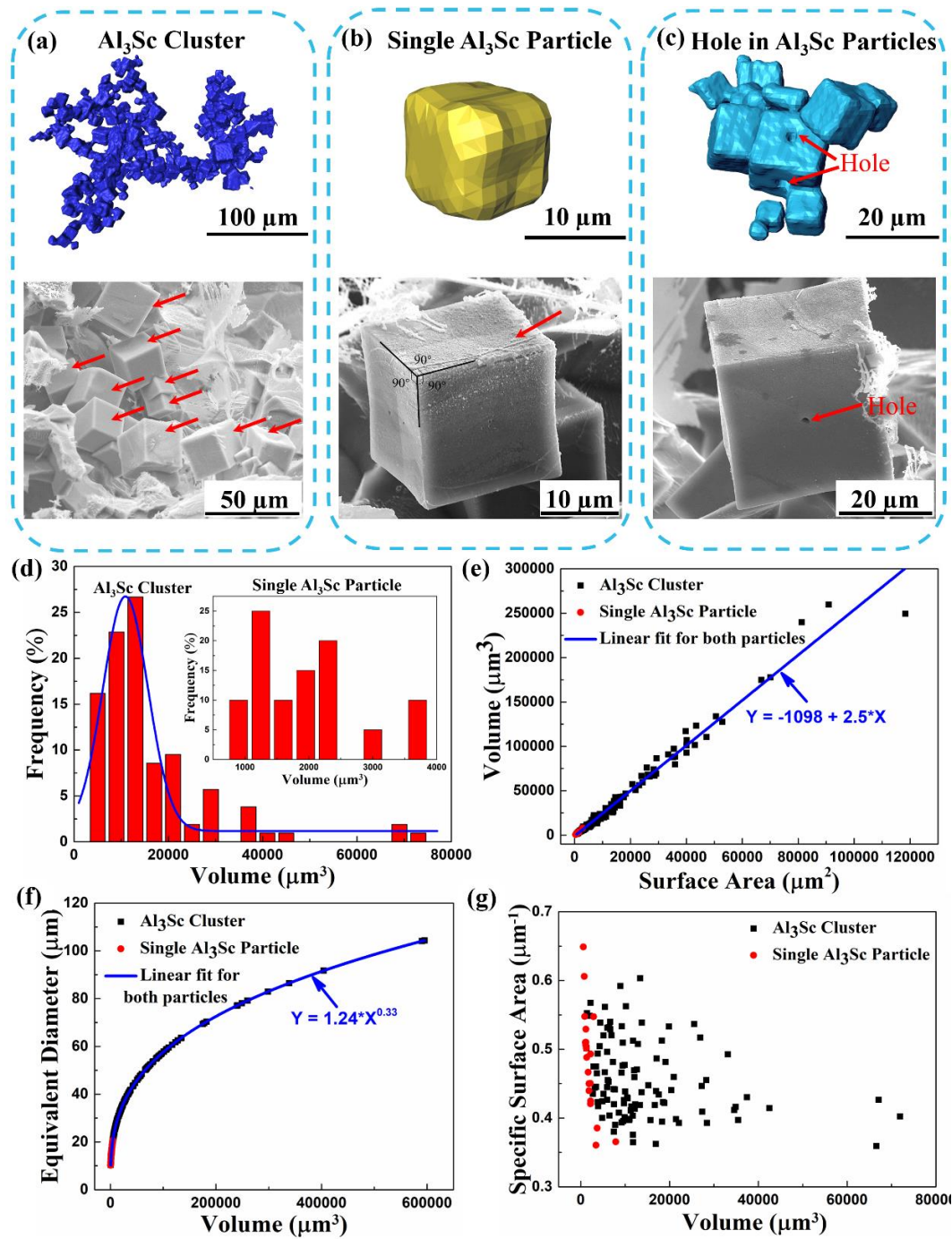

Fig. 2 3D morphologies of Al$_3$Sc phases: (a) primary Al$_3$Sc cluster; (b) single primary Al$_3$Sc particles; (c) holes in Al$_3$Sc particles; (d) volume distribution of Al$_3$Sc cluster (insert graph: single Al$_3$Sc phases ); (e) the surface area of Al$_3$Sc particles as a function of volume; (f) the volume of Al$_3$Sc particles as a function of equivalent diameter; (g) the relationship between the volume of Al$_3$Sc particles and specific surface area.

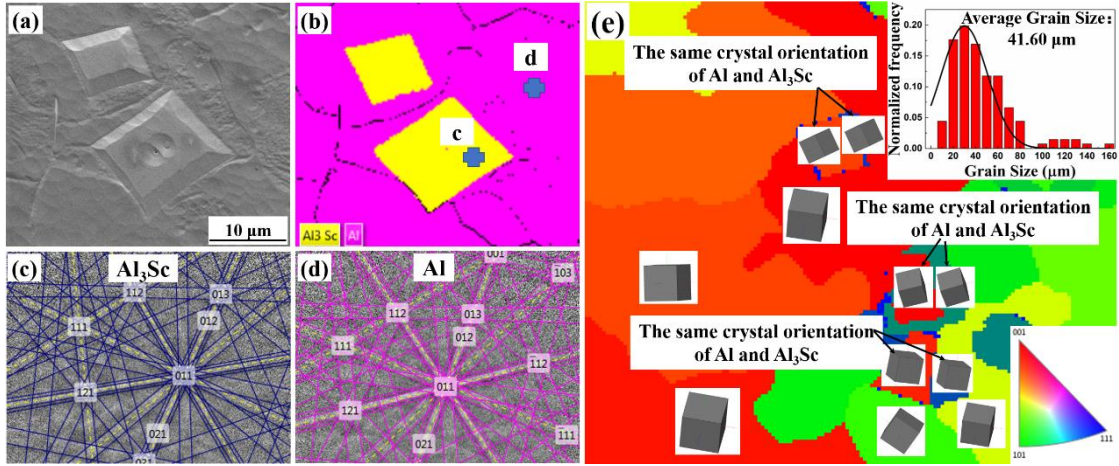

Fig. 3 EBSD analysis of Al$_3$Sc phases: (a) band contrast image; (b) EBSD image showing the phases using True Phase model; (c-d) Kikuchi patterns of Al$_3$Sc and Al phases; (e) EBSD images presents that the Al$_3$Sc in Al grains showing the same crystal orientation, insert graph shows the grain size distribution of Al matrix in the alloys.

In order to determine the possible nucleation site for Al$_3$Sc particles, FIB-TEM were performed on the sample. The detailed procedure for cutting the TEM sample can be found on the Fig. s2. Fig. 4(a) shows the region containing aluminium oxides, Al$_3$Sc particles and Al matrix, which were cut by ion beam. In order to further confirm their phase structure, the Al$_3$Sc particles/Al matrix interface (region A) and aluminium oxides/Al$_3$Sc particles interface (region B) were studied by TEM. The crystal structure of Al$_3$Sc particles and Al phase were identified by selected area diffraction pattern (SADP), as shown in Fig. 4(b). There are many tiny circle-Al$_3$Sc particles existed in the Al-side of Al$_3$Sc particles/Al matrix interface, which was confirmed by the TEM-EDS (Fig. 4(c-d)). The aluminium oxides/Al$_3$Sc particles interface was shown in the Fig. 4(e). As the TEM-EDS line mapping (Fig. 4(f-h)) indicating that the TEM results show that Al$_2$O$_3$ provides the nucleation site for Al$_3$Sc particles. The graph inserted in the Fig. 4f indentation the structure of the aluminium oxides is the α-Al$_2$O$_3$. The SEM-EDS line mapping (Fig. 4(i-j)) further confirms this result. A high density of stacking faults was observed on the Al$_3$Sc particles-side of aluminium oxides/Al$_3$Sc particles (Fig. 4k).

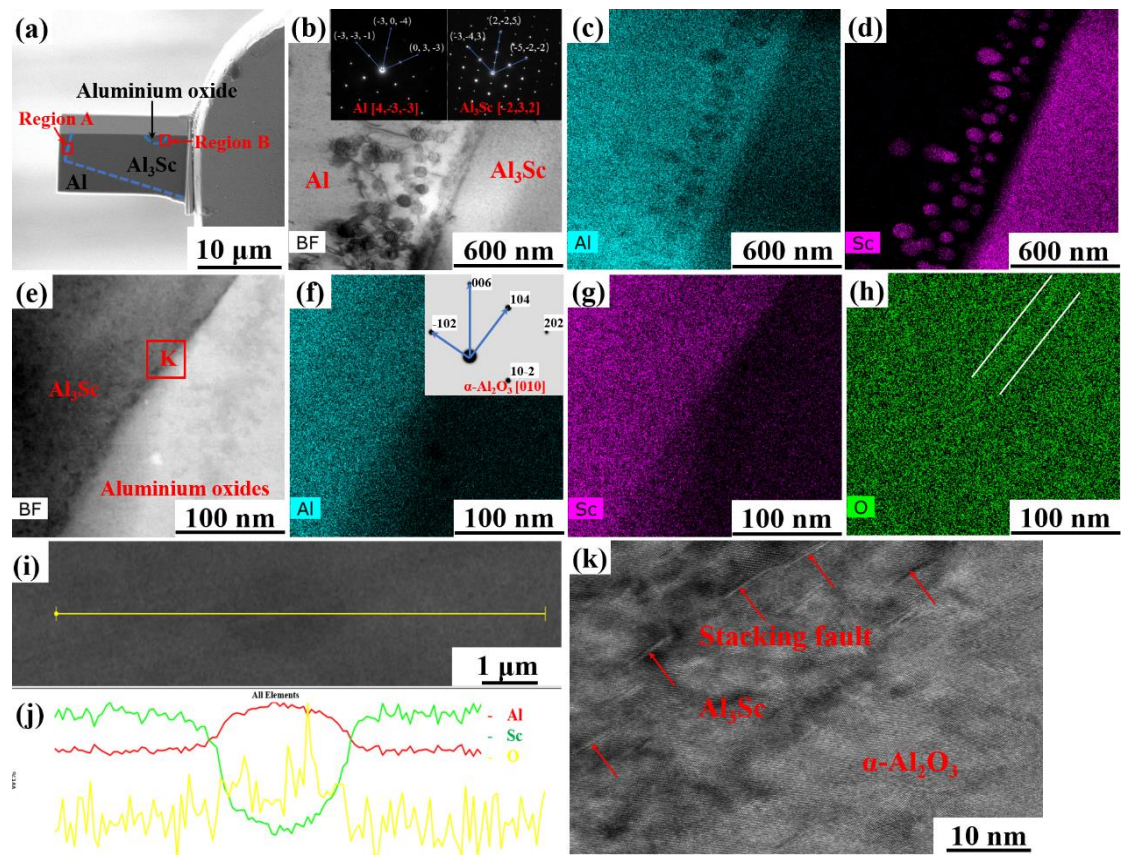

Fig. 4 FIB-TEM analysis of Al$_3$Sc phases and their interfaces: (a) the sample region contains aluminium oxides, Al$_3$Sc particles and Al matrix, which has been cut by ion beam; (b) Al$_3$Sc particles/Al matrix interface; (c-d) EDS mapping of (b); (e) aluminium oxides/Al$_3$Sc particles interface; (f-h) EDS mapping of (e); (i-j) EDS line mapping of aluminium oxides/Al$_3$Sc particles interface; (k) stacking faults on the Al$_3$Sc particles-side of α-Al$_2$O$_3$/Al$_3$Sc particles interfaces.

In this paper, the 3D structures of primary Al$_3$Sc phases in an Al-2%Sc master alloy were studied by synchrotron X-ray microtomography and electron microscopy. It was found that high mean and gaussian curvature of primary Al$_3$Sc phases are mainly at the edges and corners of the cubes. The skeletonization analysis shows that the node length between two cubes was in the range of 5-20 μm. The primary Al$_3$Sc phases are either form a cluster (volume: 5000-80000 μm$^3$), or in a single cube (volume: 1000-4000 $\mu m^3$). The surface area, equivalent diameter of $Al_3Sc$ cubes increases with increasing volume, but the specific surface area decreases. The TEM investigation confirms that α-$Al_2O_3$ are the potential nucleation site for $Al_3Sc$ phases.

**Acknowledgements**

This work was supported by UK-EPSRC grants (EP/L019965/1); the Team project of Natural Science Foundation of Guangdong Province (2015A030312003); Open Funds of National Engineering Research Centre of Near-Net-Shape Forming for Metallic Materials (2018014); Science and Technology Project of Zhaoqing (2018K009); and Scientific Research Foundation of Advanced Talents (Innovation Team), DGUT (No. KCYCXPT2016004). We also would like to acknowledge the European Synchrotron Radiation Facility, ESRF, France, for provision of synchrotron radiation beamtime on the ID19 beamline under proposal number MA 3752.